\documentclass[
aps, prb,
twocolumn,
10pt,
superscriptaddress,
]{revtex4-2}

\usepackage[T1]{fontenc}
\usepackage{lmodern}
\usepackage{amssymb}
\usepackage{amsmath}
\usepackage{graphicx}
\usepackage[dvipsnames]{xcolor}
\usepackage{siunitx}
\usepackage{microtype}
\usepackage[unicode=true]{hyperref}
\usepackage[version=4]{mhchem}
\usepackage{comment}
\usepackage{bm}
\usepackage[textsize=footnotesize]{todonotes}

\providecommand\vect[1]{\bm{#1}}

\begin{document}

\title{Defect Detection in Magnetic Systems Using U-Net and Statistical Measures}

\author{Ross Knapman}
\affiliation{Faculty of Physics and Center for Nanointegration Duisburg-Essen (CENIDE), University of Duisburg-Essen, 47057 Duisburg, Germany}
\affiliation{Institute of Mechanics, Faculty of Engineering, University of Duisburg-Essen, 45141 Essen, Germany}

\author{Atreya Majumdar}
\affiliation{Faculty of Physics and Center for Nanointegration Duisburg-Essen (CENIDE), University of Duisburg-Essen, 47057 Duisburg, Germany}

\author{Nasim Bazazzadeh}
\affiliation{Faculty of Physics and Center for Nanointegration Duisburg-Essen (CENIDE), University of Duisburg-Essen, 47057 Duisburg, Germany}
\affiliation{Radiology Department, Massachusetts General Hospital, 175 Cambridge St., Boston, MA 02114, USA}

\author{Kübra Kalkan}
\affiliation{Faculty of Physics and Center for Nanointegration Duisburg-Essen (CENIDE), University of Duisburg-Essen, 47057 Duisburg, Germany}

\author{Katharina Ollefs}
\affiliation{Faculty of Physics and Center for Nanointegration Duisburg-Essen (CENIDE), University of Duisburg-Essen, 47057 Duisburg, Germany}
\affiliation{Kirchhoff Institute for Physics, University of Heidelberg, 69120 Heidelberg, Germany}

\author{Oliver Gutfleisch}
\affiliation{Functional Materials, Institute of Materials Science, Technical University of Darmstadt, 64287 Darmstadt, Germany}

\author{Karin Everschor-Sitte}
\affiliation{Faculty of Physics and Center for Nanointegration Duisburg-Essen (CENIDE), University of Duisburg-Essen, 47057 Duisburg, Germany}

\begin{abstract}
Local material inhomogeneities can strongly influence magnetization dynamics and macroscopic magnetic properties, yet detecting such defects from magnetic imaging data remains challenging when thermal fluctuations and experimental noise obscure static contrast. Here, we investigate defect detection in strongly fluctuating magnetization regimes where signatures of inhomogeneities largely average out in time-resolved measurements. Using finite-temperature micromagnetic simulations with randomly distributed defects and material parameters representative of \ce{Ni80Fe20}, we compute per-pixel temporal mean, temporal standard deviation, and latent entropy and use them as inputs for U-Net-based semantic segmentation models. 
We find that the most effective descriptor depends on the noise level and, importantly, that robust detection requires training data that reflect the expected noise statistics. These results provide practical guidance for designing noise-robust defect-detection workflows in magnetic imaging. 
\end{abstract}

\maketitle

\section{Introduction}

Local material inhomogeneities can strongly influence magnetization dynamics in bulk and thin-film systems. Small local variations in exchange interaction, magnetic anisotropy, or easy-axis orientation can modify domain nucleation and domain-wall motion, thereby affecting macroscopic magnetic properties such as coercivity and hysteresis~\cite{Katter1996,Zheng2002,Kuhn2026,Chen2018,Yang2023,Yang2023a,Scheibel2018,Gutfleisch2016,Sepehriamin2015,lyubina2010novel}. Accurately locating and characterizing such defects is therefore important for both fundamental studies and technological applications.

Magnetic imaging techniques provide spatially resolved information on magnetization behavior, but subtle inhomogeneities can nevertheless be difficult to detect. In many situations, particularly at finite temperature or in strongly fluctuating systems, the magnetization evolves rapidly in time, so that defect-related signatures may largely average out in time-averaged magnetization images. As a result, conventional static indicators such as domain-wall width or persistent contrast features may not reliably reveal the underlying inhomogeneities. Experimental noise can further obscure weak signals in the data~\cite{Arnold1997,Kopp2015,horenko2021scalable,Rodrigues2021deeper}. Recently, machine-learning approaches, especially U-Net-style convolutional neural networks, have demonstrated promising capabilities for identifying spatial features in magnetic systems from imaging data~\cite{Cao2022,Labrie-Boulay2024,McCray2024}.

In this work, we mainly investigate defect detection in highly fluctuating regimes, where static contrast is weak, and information about defects can instead be extracted from magnetization dynamics. We develop and evaluate a combined approach that integrates statistical measures of temporal behavior with semantic segmentation. From finite-temperature micromagnetic simulations containing random defects, we compute per-pixel temporal mean, temporal standard deviation, and latent entropy - a physics-motivated measure of transition stochasticity or predictability inferred from temporally ordered state transitions~\cite{horenko2021scalable,Rodrigues2021deeper}. Each measure is used separately as input for training U-Net segmentation models. For concreteness, the simulations employ material parameters representative of \ce{Ni80Fe20}. By focusing on strongly fluctuating data, the detection task relies on robust dynamical signatures rather than on static structural features.

We evaluate segmentation performance across magnetization components and under different levels of post-processed noise applied to the simulated magnetization time series, where the added noise mimics realistic experimental acquisition conditions. 

For the magnetization dynamics considered here, we find that the temporal mean, temporal standard deviation, and latent entropy allow for defect detection at low noise levels. As the noise increases, however, we find that typically the temporal standard deviation and latent entropy emerge as the most effective measures. The temporal standard deviation primarily captures the Gaussian fluctuations of the magnetization dynamics, whereas the latent entropy is sensitive to more general stochastic transition behavior~\cite{horenko2021scalable} and therefore goes beyond purely Gaussian statistics. 
For the considered magnetization dynamics, however, this additional sensitivity plays only a minor role, leading to comparable performance of the two measures, with the temporal standard deviation performing slightly better at lower noise levels. Importantly, we also observe that achieving robustness under moderate noise requires training data with noise statistics that match those of the target data.

The remainder of the manuscript is organized as follows. In Section~\ref{sec:workflow}, we present the approach, consisting of the statistical measures and training procedures based on the synthetic data obtained from micromagnetic simulations. Sections~\ref{sec:res} and \ref{sec:dis} present the main segmentation results and discuss their behavior under varying noise conditions.

\section{Computational Workflow for Defect Detection}
\label{sec:workflow}
For automated defect detection, we followed a three-step workflow: (A) run finite-temperature micromagnetic simulations of magnetic films with randomly distributed defect regions to generate synthetic magnetization dynamics data, (B) extract scalar measures (temporal mean $\mu$, temporal standard deviation $\sigma$, and latent entropy $\mathrm{LE}$) from each pixel time series, and (C) train U-Net models to automatically segment defects from the resulting 2D measure maps.

\begin{figure*}[t]
    \centering
    \includegraphics[width=\linewidth]{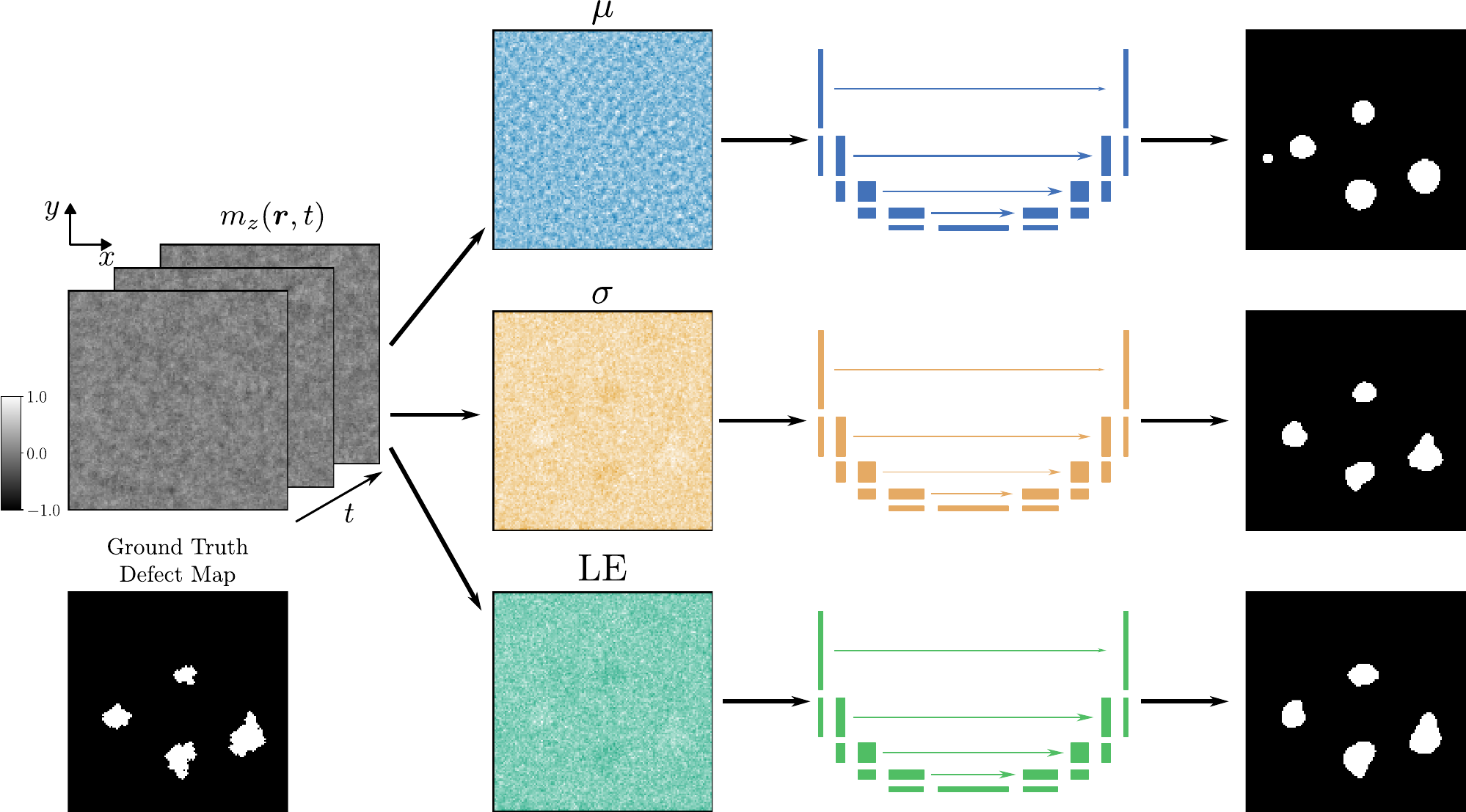}
    \caption{Schematic overview of the data processing pipeline. 
    Highly fluctuating time-series magnetization fields obtained from finite-temperature micromagnetic simulations are processed on a per-pixel basis to compute three statistical measures: the temporal mean $\mu$ (top, blue), temporal standard deviation $\sigma$ (center, yellow), and latent entropy $\mathrm{LE}$ (bottom, green). Each measure is then provided separately as a single-channel input to a U-Net trained to detect defect regions.
For the example shown, the standard deviation and latent entropy perform best.}
\label{fig:UNetSchematic}
\end{figure*}

A schematic overview of the full processing pipeline is shown in Fig.~\ref{fig:UNetSchematic}; the following subsections describe these three steps.

\subsection{Micromagnetic Simulations}

Synthetic magnetization dynamics data were generated using finite-temperature micromagnetic simulations of a thin ferromagnetic film with in-plane uniaxial anisotropy and randomly distributed impurities at approximately room temperature.
The magnetization dynamics were modeled by the stochastic Landau-Lifshitz-Gilbert equation
\begin{equation}
\frac{\mathrm{d}\vect{m}}{\mathrm{d}t} = -\gamma\,\vect{m}\times(\vect{H}_{\mathrm{eff}}+\vect{H}_{\mathrm{th}}) 
+ \alpha\,\vect{m}\times\frac{\mathrm{d}\vect{m}}{\mathrm{d}t},
\end{equation}
where $\vect{m}(\vect{r},t)$ is the unit magnetization vector with Cartesian components $(m_x,m_y,m_z)$, $\gamma$ is the gyromagnetic ratio, $\alpha$ the Gilbert damping parameter, and $\vect{H}_{\mathrm{th}}$ a Langevin thermal field representing fluctuations at temperature $T$~\cite{Brown1963, Lyberatos1993, Leliaert2017}. In this work, we report results for the in-plane component $m_x$ and the out-of-plane component $m_z$. The component $m_y$ is omitted due to the normalization condition $|\vect{m}|=1$.
The effective magnetic field $\vect{H}_{\mathrm{eff}}=-(\mu_0 M_\mathrm{s})^{-1}\,\delta E/\delta \vect{m}$ is obtained from a micromagnetic energy density functional with spatially varying exchange and uniaxial anisotropy, allowing us to distinguish between the background material and localized defect regions,
\begin{equation}
\begin{aligned}
E &= \int \! \Big[
A(\vect{r})(\nabla \vect{m})^2
+ K(\vect{r}) \big(1-(\vect{m}\cdot\hat{\vect{u}}(\vect{r}))^2\big) \\
&\quad - \frac{\mu_0}{2} M_\mathrm{s}\,\vect{m}\cdot \vect{H}_\mathrm{d}
\Big]\, \mathrm{d}V,
\end{aligned}
\end{equation}
where $A(\vect{r})$ is the exchange stiffness, $K(\vect{r})$ the uniaxial anisotropy constant, $\hat{\vect{u}}(\vect{r})$ the local easy-axis direction, and $\vect{H}_\mathrm{d}$ the demagnetizing field.
Simulations were performed on a $128\times128\times1$ grid with cell dimensions $(\SI{2}{\nano\meter},\SI{2}{\nano\meter},\SI{10}{\nano\meter})$ using \texttt{MuMax3}~\cite{Vansteenkiste2014}.

Background material parameters were set to values representative of \ce{Ni80Fe20}: $A=\SI{1.0e-11}{\joule\per\meter}$, $K=\SI{1.5e2}{\joule\per\meter\cubed}$, and everywhere, $M_\mathrm{s}=\SI{8.4e5}{\ampere\per\meter}$, damping $\alpha=0.01$~\cite{Coey2010}, $T=\SI{300}{\kelvin}$, with the background easy axis aligned along the $x$ direction, i.e.~$\hat{\vect{u}}=\hat{\vect{x}}$. 

Four non-overlapping inhomogeneity regions were placed at random positions for each simulation run. Each inhomogeneity was generated on the $128\times128$ grid as an irregular, simply connected blob with a target size on the order of a few cells. 
Within inhomogeneity regions, exchange stiffness and anisotropy were scaled relative to the background with fixed factors: two inhomogeneity regions were set to $1.2$ times the background value and two to $0.8$ times the background value in every simulation (i.e., this scaling pattern was not randomized between runs). The in-plane easy-axis orientation was randomly perturbed around $\hat{\vect{x}}$ by drawing angles from a Gaussian distribution with a \SI{20}{\degree} standard deviation, while the background anisotropy remained along $\hat{\vect{x}}$.

The full dataset consisted of 2048 independent simulations, with ground-truth defect masks constructed directly from the known defect regions in each simulation. 
Initial magnetization was uniformly aligned along the easy axis $x$. Each simulation extended over \SI{1}{\nano\second} with the magnetization output every \SI{10}{\pico\second} (100 frames).
From each simulation, we then derived per-pixel summary measures from the magnetization time series.

\subsection{Measures and Latent Entropy}

To capture different aspects of local dynamics, we computed per-pixel summary measures from magnetization time series: temporal mean $\mu$, temporal standard deviation $\sigma$, and latent entropy $\mathrm{LE}$~\cite{horenko2021scalable,Rodrigues2021deeper}. To assess robustness to experimental noise, we synthetically perturbed each magnetization-component time series (here $m_x$ or $m_z$) before computing the measures. We considered two standard zero-mean noise models with variance $\sigma_n^2$: additive Gaussian noise,
\begin{equation}
\tilde{m}_i = m_i + n, \qquad n \sim \mathcal{N}(0,\sigma_n^2),
\end{equation}
and multiplicative speckle noise,
\begin{equation}
\tilde{m}_i = m_i(1+n), \qquad n \sim \mathcal{N}(0,\sigma_n^2).
\end{equation}
The measures $\mu$, $\sigma$, and $\mathrm{LE}$ were then recomputed from the noisy series.
The latent entropy~\cite{horenko2021scalable} was computed for each pixel time series ($m_x$ or $m_z$). The signal was discretized into $n_\mathrm{bins}=5$ states, and the latent entropy was evaluated following the procedure described in Ref.~\cite{horenko2021scalable}. 

\subsection{U-Net Architecture and Model Training}

For semantic segmentation of defect regions, we used a U-Net architecture~\cite{ronneberger2015u} trained on single-channel measure maps (\(\mu\), \(\sigma\), or \(\mathrm{LE}\)) for each magnetization component (\(m_x\), \(m_z\)). Separate models were trained for each measure/component combination, and robustness was tested using both clean and noise-matched training conditions. Full details of the architecture, optimization setup, data splitting, normalization, and model selection are provided in Appendix~\ref{app:unet-training}.

\section{Results}
\label{sec:res}

\begin{figure*}
    \centering
    \includegraphics[width=\linewidth]{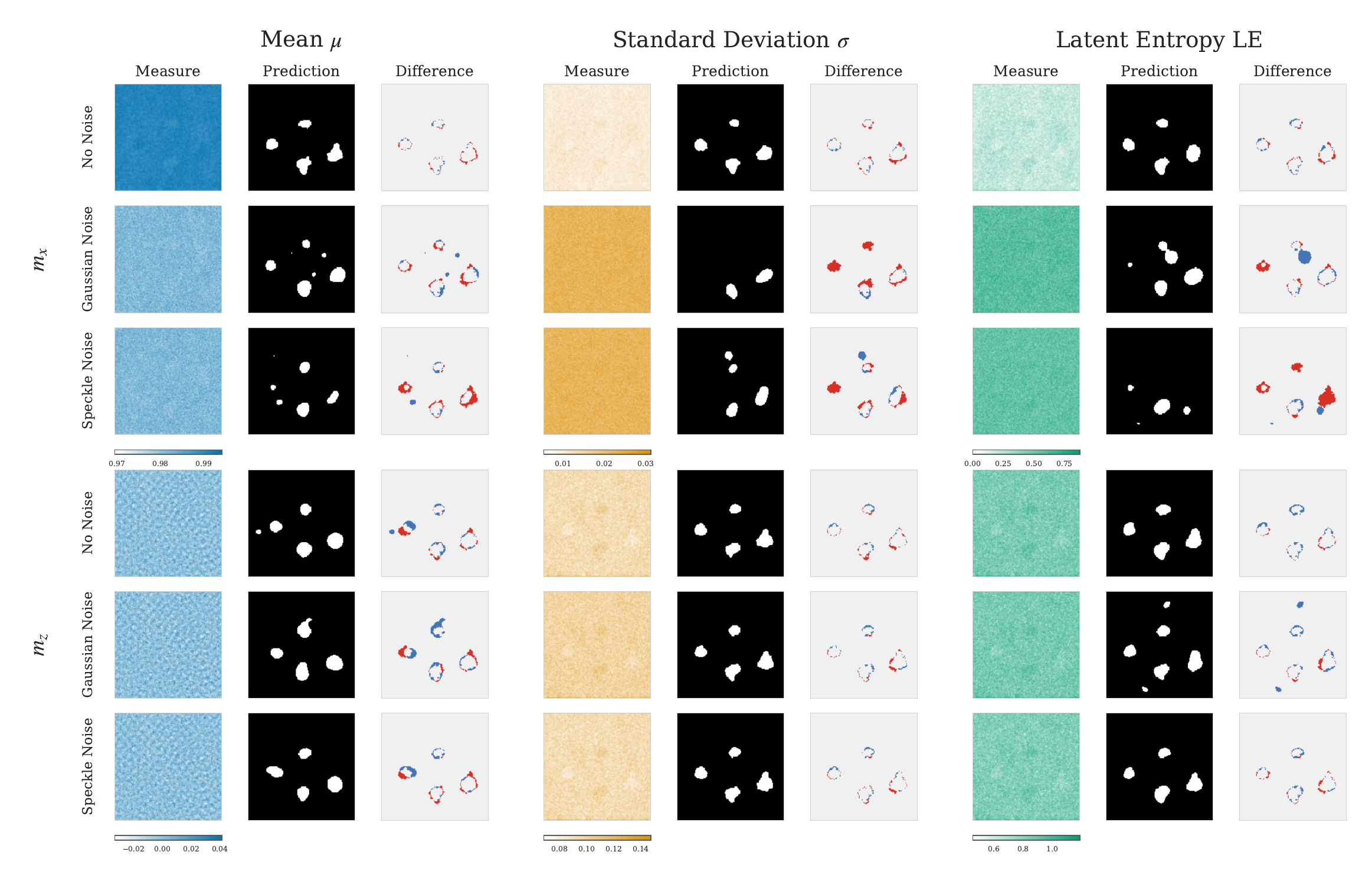}
    \caption{
    Qualitative prediction overview for defect segmentation using the three input measures: temporal mean ($\mu$), temporal standard deviation ($\sigma$), and latent entropy (LE). Columns are grouped by measure, and within each group show (from left to right) the input measure map, the U-Net prediction, and the prediction difference relative to the ground truth shown in Fig.~\ref{fig:UNetSchematic} (false positives/negatives highlighted in blue/red respectively). Rows are organized by magnetization component ($m_x$ top block, $m_z$ bottom block) and noise condition (no noise, additive Gaussian noise, and multiplicative speckle noise). Noisy examples are evaluated using models trained with matching noise statistics.
    }
    \label{fig:PredictionsWithNoise}
\end{figure*}

\begin{figure*}
    \centering\includegraphics[width=\linewidth]{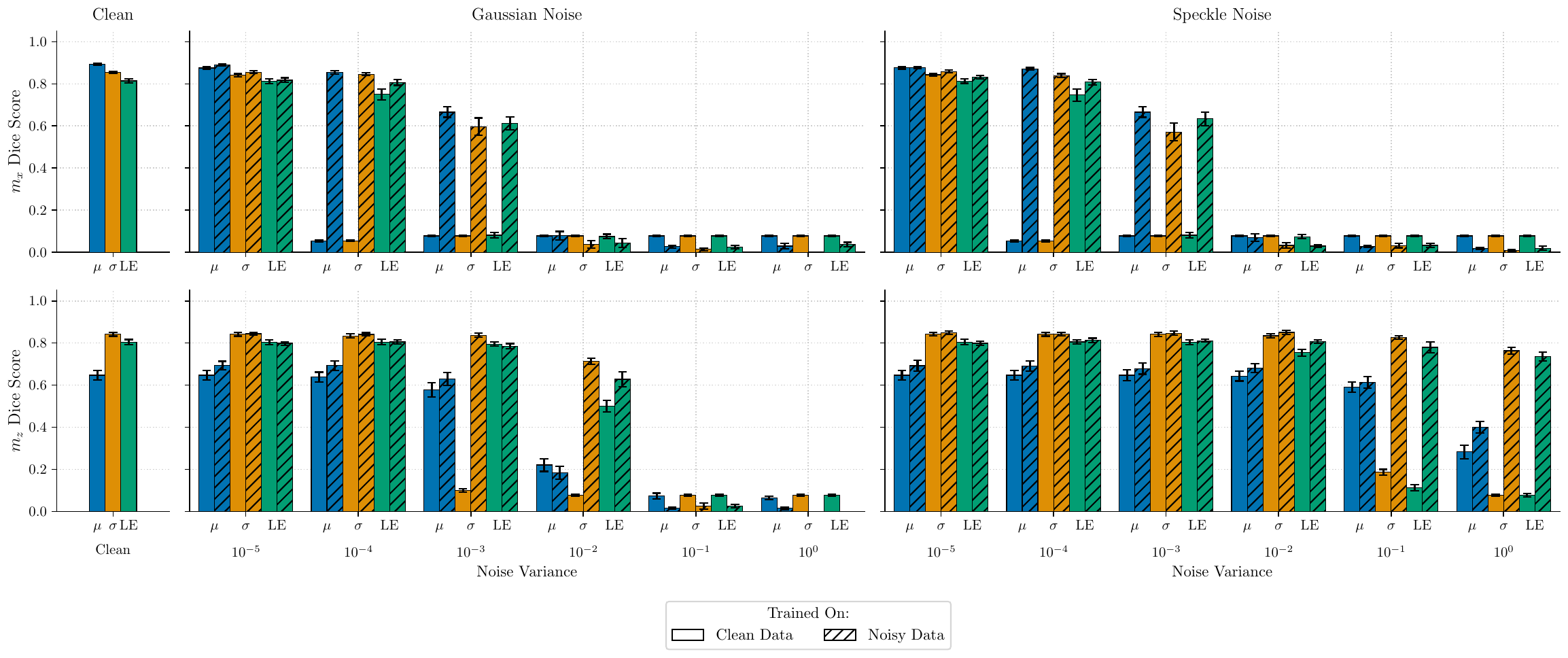}
    \caption{
        Test-set Dice coefficients for defect segmentation from the three single-channel inputs ($\mu$, $\sigma$, LE).
Top row: $m_x$; bottom row: $m_z$.
The left column reports clean-data performance, while the center and right columns show performance versus noise variance for additive Gaussian and multiplicative speckle noise, respectively.
For each measure and noise level, results are shown for models trained on clean data and for models trained with matching noisy data, as indicated in the legend.
}
    \label{fig:DiceScoreComparison}
\end{figure*}

We evaluated the ability of a U-Net to segment defect regions from single-channel inputs derived from the magnetization time series on the held-out test set.
Representative predictions are shown in Fig.~\ref{fig:PredictionsWithNoise}
and quantitative performance is summarized in Fig.\ref{fig:DiceScoreComparison}, which compares models trained on clean data with models trained using matching noisy data under clean, additive-Gaussian, and multiplicative-speckle evaluation conditions.

\subsection{Baseline Performance on Clean Data}

On clean (noise-free) inputs, Fig.~\ref{fig:PredictionsWithNoise} already shows a strong component dependence in visual contrast. For $m_x$, defect regions are visible in the mean map, while for $m_z$ the mean map shows much weaker visual contrast between defect and background regions. This qualitative difference is consistent with the quantitative Dice scores in Fig.~\ref{fig:DiceScoreComparison}. For the in-plane component $m_x$, the temporal mean provides the strongest single-channel signal (baseline Dice $\approx 0.89$), followed by the temporal standard deviation ($\approx 0.85$) and latent entropy ($\approx 0.81$). For the out-of-plane component $m_z$, the temporal mean is markedly less informative (baseline Dice $\approx 0.65$), whereas the temporal standard deviation ($\approx 0.84$) and latent entropy ($\approx 0.80$) remain competitive.

Across all measures, the predicted masks in Fig.~\ref{fig:PredictionsWithNoise} also tend to have smoother, more rounded boundaries than the jagged pixel-level contours in the ground-truth masks. This can be partially explained by the static magnetization having a characteristic length scale $\sqrt{A/K}$ which results in smooth transitions between the defect regions and background. Furthermore, this behavior is expected for U-Net-style segmentation: the encoder-decoder pathway with pooling and upsampling favors spatially coherent regions, and optimization of pixel-wise cross-entropy and Dice overlap rewards correct area overlap more than reproducing high-frequency boundary roughness. After thresholding soft output probabilities, small one-pixel protrusions in the labels are therefore often suppressed, leading to slightly smoother predicted defect contours~\cite{Zunair2021,Kamath2025}.

\subsection{Robustness Under Additive (Gaussian) Noise}

Under weak Gaussian noise (variance $10^{-5}$), Dice scores remain close to their clean baselines for all measures (Fig.~\ref{fig:DiceScoreComparison}), and noise-augmented training provides only marginal improvements. At moderate noise levels (variance $10^{-4}$ to $10^{-3}$), the benefit of matched noise augmentation becomes pronounced: models trained only on clean data fail (Dice $\lesssim 0.1$ for several measures), whereas noise-trained models retain high accuracy. For example, for $m_x$ at variance $10^{-4}$, noise-trained models achieve Dice $\approx 0.85$ (mean) and $\approx 0.85$ (standard deviation), compared with $\approx 0.05$ for the corresponding clean-trained models. For $m_z$, the standard-deviation input is particularly robust when trained with matching noise (Dice $\approx 0.84$ at variance $10^{-4}$ and $\approx 0.84$ at variance $10^{-3}$). Latent entropy benefits similarly from noise-aware training at these noise levels, but remains below the best-performing measure for each component.

At large Gaussian noise variance ($\gtrsim 10^{-2}$), performance degrades substantially even for noise-trained models, with Dice scores dropping below $\sim 0.1$ for most measure/component combinations. Representative noisy examples at variance $10^{-3}$ are shown in Fig.~\ref{fig:PredictionsWithNoise}, where the input measures are strongly distorted and the predicted masks show missed defects and false positives, consistent with the sharp deterioration in quantitative scores.
At the highest noise levels, the two training strategies fail in qualitatively different ways, which explains the non-zero Dice values visible in Fig.~\ref{fig:DiceScoreComparison} for several clean-trained results (for both Gaussian and speckle noise). Clean-trained models often collapse to near-all-defect predictions (the defect class assigned to most pixels). Because the true defect mask occupies only a limited but non-zero area, this near-constant prediction can still yield a finite Dice overlap. In contrast, models trained with noisy data at these extreme noise levels more often collapse to near-all-non-defect predictions (defects mostly missed), for which the Dice score tends toward zero. Thus, the residual Dice of clean-trained models at very high noise should not be interpreted as better segmentation quality, but as a different failure mode.

\subsection{Robustness Under Multiplicative (Speckle) Noise}

The qualitative trends under speckle noise are similar at low-to-moderate variance, with matched noise augmentation again being critical once the noise level exceeds the weak-noise regime (Fig.~\ref{fig:DiceScoreComparison}). However, compared to additive Gaussian noise, multiplicative speckle noise is better tolerated for certain inputs and components. In particular, for $m_z$ the standard-deviation and latent-entropy inputs remain robust over a wider range of speckle variances when trained with matching noise (Dice remaining $\gtrsim 0.73$ even at variance $1.0$), whereas $m_x$ performance drops more rapidly at high speckle variance. This component dependence suggests that the out-of-plane fluctuations captured by $m_z$ contain a comparatively stable signature of inhomogeneity regions under multiplicative intensity distortions.
Within the high-variance tail for $m_x$, latent entropy can also edge out the mean and standard-deviation inputs (e.g., speckle variance $0.1$ and $1.0$), though all measures are weak in that regime.



\section{Discussion}
\label{sec:dis}

The results highlight three main messages for defect detection in magnetic imaging. 
First, the most informative input measure depends on the specific magnetization dynamics and the associated fluctuations, which are determined by the physical model and may manifest differently across magnetization components. For the dynamics considered here, the temporal mean of $m_x$ produced the strongest defect contrast, whereas for $m_z$ the temporal standard deviation was more informative.
This indicates that feature design should be selected for the specific signal channel and contrast mechanism, rather than assumed to transfer unchanged between magnetization components.

Second, robustness is determined primarily by how well training data match inference conditions. Models trained only on clean inputs performed well on clean test data, but degraded strongly under moderate noise. In contrast, training with matched noise preserved high Dice scores over a broad range of Gaussian and speckle perturbations. For practical deployment, this supports two complementary strategies: incorporating realistic noise during training and updating pretrained models with domain-adaptation or transfer-learning steps when measurement conditions shift.
One possible interpretation of the particularly fast drop for clean-trained $m_x$ models is that they may rely more heavily on high-contrast intensity cues that are prominent in clean $m_x$ maps, whereas $m_z$ models are forced to use subtler structure already in the clean case. Under added noise, those dominant $m_x$ cues may be disrupted first, causing a sharper loss in segmentation quality.

Third, latent-entropy maps are a useful complementary descriptor. They were competitive with standard deviation maps on clean data and remained particularly robust for $m_z$ under speckle noise, but they were not consistently superior across all regimes for the magnetization dynamics considered here.

For experimental applications, we suggest a simple guideline. The temporal mean can be preferred when strong in-plane--like contrast is present (analogous to \(m_x\)), whereas the temporal standard deviation or latent entropy are more suitable when the out-of-plane, like contrast is weaker (analogous to \(m_z\)). As a default noise model, we recommend additive Gaussian noise, since additive-noise descriptions are commonly employed in magnetic imaging acquisition and denoising workflows, where detector-related noise often represents a key practical limitation~\cite{Arnold1997,Kopp2015}. Rather than relying on training with clean data only, the noise level should be estimated from background regions, and the model should be trained or fine-tuned using data with matching noise statistics, for example, obtained from a short calibration dataset.

Beyond these central findings, this work provides a controlled benchmark that can be extended systematically. We deliberately used synthetic defect distributions, a single geometry, and a restricted range of defect morphologies and parameter variations to enable reproducible comparisons across measures and noise models. Likewise, the single-channel training design isolated the contribution of each descriptor. A natural next step is multi-channel segmentation that combines mean, standard deviation, and latent entropy as joint inputs, which may further improve accuracy and robustness.

These results provide practical guidance for designing noise-robust defect-detection workflows in magnetic imaging and illustrate how dynamical magnetization signatures can be leveraged for reliable identification of material inhomogeneities.


\section{Data and Code Availability}

All code used to generate the synthetic data, compute the measures, and train the U-Net models will be released in a public repository upon publication; the persistent archive DOI will be added in the published version.

\section{Acknowledgments}
We thank Illia Horenko, Omer Fetai, Philipp Gessler, and Heiko Wende for discussions. We acknowledge funding from the German Research Foundation (DFG) Project-ID 405553726 (CRC/TRR 270, projects B12, A03 and B01) and Project No. 505561633 in the TOROID project cofunded by the French National Research Agency ANR under Contract No. ANR-22-CE92-0032.

AI-assisted language editing was used.

\appendix

\section{U-Net Architecture and Model Training}
\label{app:unet-training}

\begin{figure*}
    \centering
    \includegraphics[width=\linewidth]{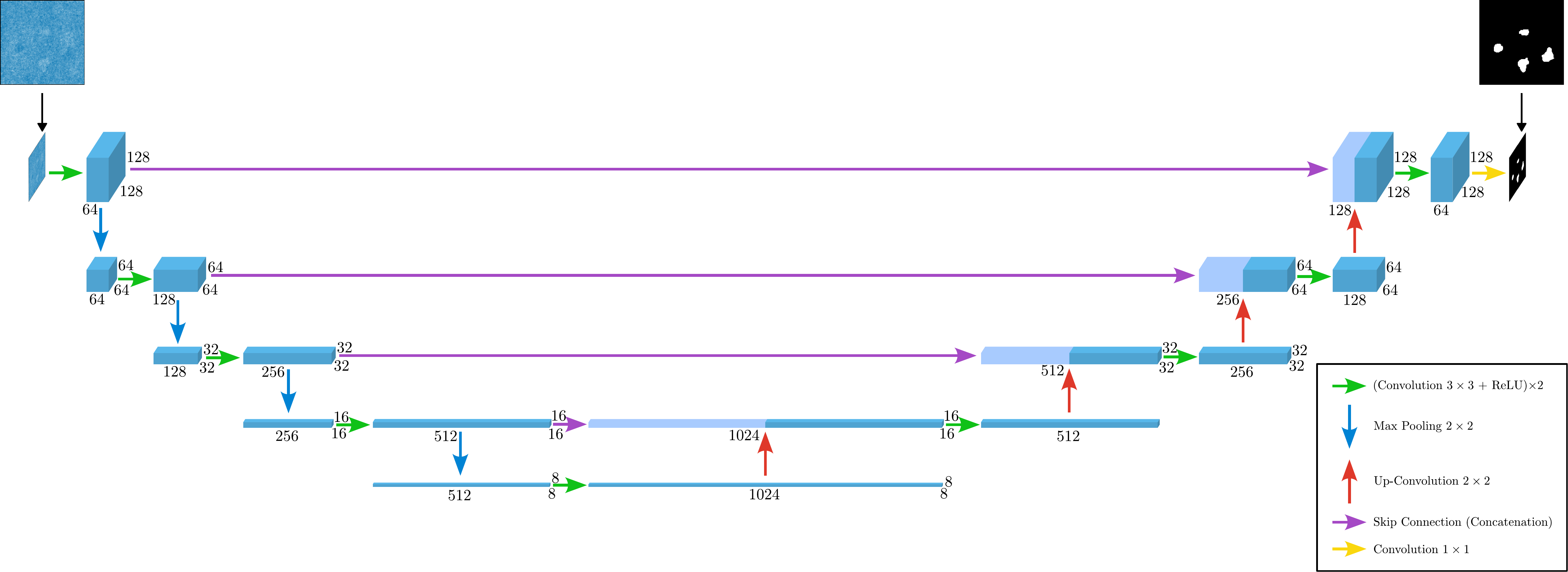}
    \caption{
    U-Net architecture used for defect segmentation.
    The network consists of a four-level encoder--decoder structure with feature sizes 64--128--256--512 and a 1024-channel bottleneck.
    Each block contains two $3 \times 3$ convolutional layers with ReLU activation.
    Downsampling is performed using $2 \times 2$ max pooling, and upsampling using $2 \times 2$ transposed convolutions.
    Skip connections concatenate encoder feature maps with the corresponding decoder levels.
    A final $1 \times 1$ convolution maps the features to two output channels (defect and non-defect logits).
    Numbers indicate feature-channel counts at each resolution level.
    }
    \label{fig:U-NetArchitecture}
\end{figure*}

For pixel-level segmentation of defect regions, we employed a U-Net convolutional neural network architecture, originally developed for biomedical image segmentation~\cite{ronneberger2015u}. The network comprises an encoder-decoder structure with skip connections between corresponding layers. Our implementation uses four resolution levels with feature sizes 64--128--256--512 and a bottleneck of 1024 channels. Each block consists of two $3\times 3$ convolutions with ReLU activations; downsampling is performed by $2\times 2$ max pooling and upsampling by transposed convolutions. The architecture is illustrated in Fig.~\ref{fig:U-NetArchitecture}.

Each model received a single input channel (one measure map per magnetization component) and output two channels representing defect and non-defect logits. We trained separate models for each combination of measure $\{\mu,\sigma,\mathrm{LE}\}$ and component $\{m_x,m_z\}$.

Training used the Adam optimizer with a learning rate of $10^{-4}$, batch size 32, and 500 epochs. Dropout regularization (spatial/channel-wise) with probability $p=0.1$ was inserted inside each double-convolution block after the first Conv+ReLU and before the second Conv+ReLU; this was applied in all encoder blocks, the bottleneck, and all decoder blocks. The training conditions were chosen after having tested different hyperparameters and regularization methods. The dataset was split into training/validation/test subsets in a 0.70/0.15/0.15 ratio, ensuring reproducible splits across experiments.

Inputs were normalized separately for each measure/component combination using the mean and standard deviation computed on the noise-free training subset; these statistics were then applied unchanged to all noise conditions. The loss function was the pixel-wise cross-entropy between predicted logits and the binary ground-truth labels. Model performance was assessed using the validation Dice coefficient, and the checkpoint with the best validation Dice score was used for all test-set evaluations.

Model performance is quantified by the Dice coefficient, which ranges from 0 (no overlap between predicted and ground-truth masks) to 1 (perfect overlap). This configuration enables the network to learn spatial signatures of magnetic inhomogeneities from both clean and noise-degraded inputs.

\sloppy
\bibliographystyle{apsrev4-2}
\bibliography{References}

\end{document}